\documentclass[
reprint,
longbibliography,
amsmath, 
amssymb, 
aps, 
superscriptaddress,
pra,
]
{revtex4-2}
\usepackage{graphicx}% Include files
\usepackage{dcolumn}% Align table columns on decimal point

\usepackage{bm}% bold math
\usepackage{braket} %For the representation of bra and ket
\usepackage{hyperref}% add hypertext capabilities
\usepackage{subfigure} % subgraficos
\usepackage{nccmath}
\usepackage{amsthm} % Theorem Formatting
\usepackage{mathrsfs} %for the Hilbert symbol
\usepackage{xfrac} %for the fraction 1/2
\usepackage{pifont}
\usepackage[export]{adjustbox}
\usepackage{bbm} %probability font
\usepackage[utf8]{inputenc} 
\usepackage[normalem]{ulem}
\usepackage[usenames,dvipsnames,svgnames,table]{xcolor}
\usepackage[sort&compress]{natbib}
\usepackage{color}

\usepackage{longtable}
 
\usepackage{commath}
\usepackage{multirow}

\newcommand{\Tr}{{\rm Tr}}

\begin{document}

\title{Diagnosing crosstalk in large-scale QPUs using zero-entropy classical shadows}

\author{J. A. Monta\~nez-Barrera}
 	\altaffiliation{Corresponding author: J. A. Monta\~nez-Barrera; j.montanez-barrera@fz-juelich.de}
	\affiliation{Jülich Supercomputing Centre, Institute for Advanced Simulation, Forschungszentrum Jülich, 52425 Jülich, Germany}
\author{G. P. Beretta}
	\affiliation{Mechanical and Industrial Engineering Department, Universit$\grave{a}$ di Brescia, via Branze 38, 25123 Brescia, Italy}
 \author{Kristel Michielsen}
	\affiliation{Jülich Supercomputing Centre, Institute for Advanced Simulation, Forschungszentrum Jülich, 52425 Jülich, Germany}
	\affiliation{AIDAS, 52425 Jülich, Germany}
	\affiliation{RWTH Aachen University, 52056 Aachen, Germany}
\author{Michael R. von Spakovsky}
	\altaffiliation{vonspako@vt.edu (M.R. von Spakovsky)}
	\affiliation{Department of Mechanical Engineering, Virginia Tech, Blacksburg, VA 24061, USA}

\begin{abstract}

As quantum processing units (QPUs) scale toward hundreds of qubits, diagnosing noise-induced correlations (crosstalk) becomes critical for reliable quantum computation. In this work, we introduce Zero-Entropy Classical Shadows (ZECS), a diagnostic tool that uses information of a rank-one quantum state tomography (QST) reconstruction from classical shadow (CS) information to make a crosstalk diagnosis. We use ZECS  on trapped ion and superconductive QPUs including ionq\_forte (36 qubits), ibm\_brisbane (127 qubits), and ibm\_fez (156 qubits), using from 1,000 to 6,000 samples. With these samples, we use the ZECS to characterize crosstalk among disjoint qubit subsets across the full hardware. This information is then used to select low-crosstalk qubit subsets on ibm\_fez for executing the Quantum Approximate Optimization Algorithm (QAOA) on a 20-qubit problem. Compared to the best qubit selection via Qiskit transpilation, our method improves solution quality by 10\% and increases algorithmic coherence by 33\%. ZECS offers a scalable and measurement-efficient approach to diagnosing crosstalk in large-scale QPUs.

\begin{description}
	\vspace{0.2cm}
	\item[Keywords] zero-entropy classical shadow, quantum state tomography, routing qubits, IBM Quantum, 
    QPU crosstalk, IonQ Forte.
\end{description}

\end{abstract}

\maketitle
\section{Introduction}

\begin{figure*}[t]
\centering
\includegraphics[width=18cm]{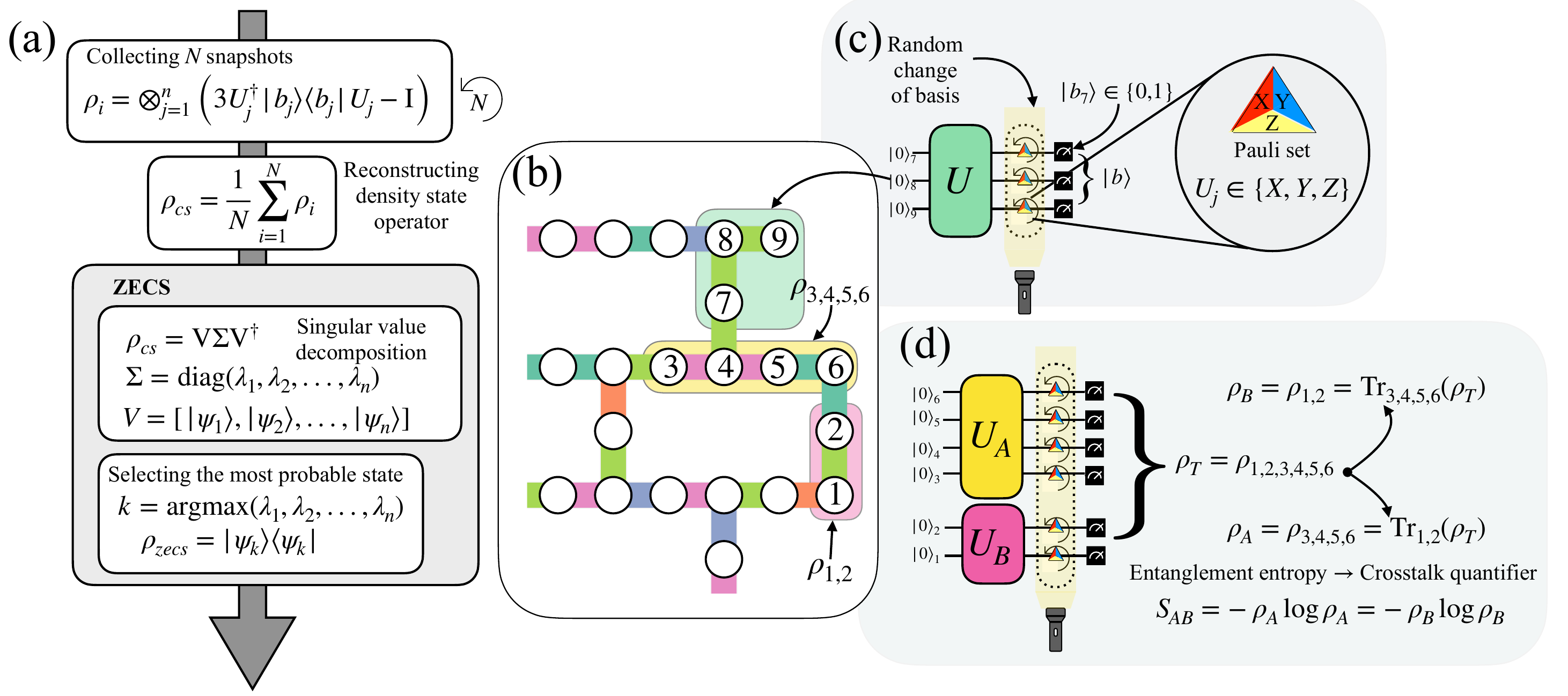}
\caption{\label{Fig:CS} (a) Shows the ZECS density state operator reconstruction workflow. (b) Shows the QPU layout with three sets of qubits chosen to implement some circuits. Vertices represent qubits and edges two-qubit native interactions of the QPU. (c) Shows the acquisition of the CS information to reconstruct the density state operator. After implementing the circuit, the measuring basis changes randomly from the Pauli set, and the process repeats $N$ times to get $N$ snapshots. (d) One ZECS application is the quantification of the entanglement entropy from disjoint circuits. First, the full-density state operator $\rho_T$ is reconstructed, and then the partial trace to the disjoint circuits is applied. If there is no leakage of information, the entanglement entropy $S_{AB} \approx  0$.}
\end{figure*}

In recent years, the capabilities of quantum processing units (QPUs) have grown to the size of hundreds of qubits \cite{Bluvstein2024, Kim2023}. At this scale, sources of noise only visible at large circuit sizes, such as crosstalk, are emerging. Crosstalk, understood here as a platform-independent phenomenon, refers to errors in the execution of quantum circuits that originate from interactions with the immediate neighborhood of the targeted qubits and beyond. These include correlated errors between distant qubits and gate errors induced by qubits not directly involved in the gate operations \cite{Sarovar_2020}. Different studies have investigated the nature and characterization of crosstalk, including \cite{Gambetta_2012, Piltz_2014, Sheldon_2016, Rudinger_2019, Sarovar_2020, Debroy_2020}.  

Usually, characterization methods show how noise affects individual qubits, 2-qubit gates, or quantum circuits. Different methods have been proposed for the characterization of quantum hardware, e.g., quantum state tomography \cite{DAriano2002} (QST), randomized benchmarking (RB)\cite{Emerson2005}, quantum volume (QV) \cite{Moll2018}, cross-entropy benchmarking (XEB)\cite{Boixo2018}, algorithmic qubits (AQ)\cite{Chen2023}, or error per layered gate (EPLG)\cite{McKay2023}.

From these techniques, one that recovers detailed information about a quantum system is QST. It searches through all the degrees of freedom of the density operator $\rho$, which for $n_q$ qubits scales as $2^{2n_q + 1}$. In practice, common methods to calculate the QST of $\rho$ for a sampling error $\varepsilon$ require at least $O(4^{n_q}/\varepsilon^2)$ copies of $\rho$ \cite{Haah2017, Gebhart2023} with a lower bound of $O(3^{n_q}/\varepsilon^2)$ \cite{chen2024optimaltradeoffentanglementcopy}. As a result, calculating the density operator becomes computationally impractical, even for systems with only a few qubits on real quantum hardware.

In this work, we propose a new methodology to characterize QPUs performance and crosstalk. To this end, we developed zero-entropy classical shadow (ZECS), a methodology that uses classical shadows (CS) \cite{Huang2020} as a subroutine to reconstruct the density state operator of small QPUs subsystems. 

CS tool has been used to get information on the expected values of quantum systems. For example, in \cite{Huang2022}, CS is used to predict the ground state properties of many-body physics problems, combining CS with a polynomial-time machine learning algorithm. It has also been applied to predicting quantum Fisher information \cite{Rath2021}, quantum process tomography \cite{Levy2024}, and information scrambling \cite{McGinley2022}. The effects of noise on classical shadows have been analyzed in \cite{Koh2022}, and methods to mitigate these effects, including real QPU implementations, have been demonstrated in \cite{Jnane2024, McGinley2022, Levy2024}.

 Figure \ref{Fig:CS}(a) shows the workflow of the ZECS methodology. First, classical snapshots of the density state operator are collected using random single-qubit Clifford circuits. Then, the density state operator $\rho_{cs}$ of these sections is reconstructed using the mean value of the different snapshots. Up to this point, the reconstruction is unstable, and $\rho_{cs}$ is not positive semidefinite, a needed condition of the density state operator. To correct this, we use the rank-one reconstruction of the density state operator, which we denominate Zero Entropy CS (ZECS). This methodology projects $\rho_{cs}$ into the pure state of the eigenvector associated with the largest eigenvalue of $\rho_{cs}$, artificially removing the entropy associated with it. This strategy makes the density state operator generated, $\rho_{zecs}$, to describe a positive semidefinite and unit trace matrix.

Since the number of samples required to reconstruct density state operators is expected to grow exponentially with the number of qubits, as in other QST methods, ZECS is employed to gain insights into the behavior of small sections of quantum processing units (QPUs) even if the qubits involved in the reconstruction are spatially apart.

The benefit of ZECS comes from its reusability: a single batch of measurements can be reused to estimate an arbitrary density state operator of any set of qubits from the QPU, requiring at most $O(d \log^2 d))$ random settings, where $d=2^{n_q}$, and assuming a rank-one (pure state) reconstruction \cite{Gross_2010}. Therefore, for a trace distance error, $\varepsilon$, the number of samples required is of $O(n_q^2 * 2^{n_q}/\varepsilon^2)$.  In contrast, conventional QST requires $O(3^{n_q}/\varepsilon^2)$ samples for each pair of qubits involved in the protocol. While ZECS provides less resolution to  distinguish different noise processes, we show that it is well-suited for detecting crosstalk.

Figure~\ref{Fig:CS}(b) shows a section of ibm\_brisbane layout, where 3 disjoint circuits are used (1,2), (3,4,5,6), and (7,8,9). The ZECS methodology is flexible enough to reconstruct the 3 disjoint sections or combinations of them. Fig. \ref{Fig:CS}(c) shows the reconstruction of one of the disjoint sections, qubits (7,8,9). Using $\rho^{zecs}_{7,8,9}$, a diagnostic can be done in terms of, for instance, the fidelity (F) or trace distance (D). 

Fig. \ref{Fig:CS}(d) shows another application of ZECS. It is based on the reconstruction of the density state operator involving qubits (1,2,3,4,5,6), $\rho_T$, which has the disjoint circuit $\rho_B \rightarrow$ (1,2) and $\rho_A \rightarrow$ (3,4,5,6). What is expected is that the entanglement entropy $S_{AB}$ is close to zero. A large value of $S_{AB}$, on the other hand, would indicate that crosstalk between the qubits is involved.

Experimentally, ZECS is used to make a diagnostic in terms of $F$, (or infidelity, $1-F$), $D$, and the entanglement entropy ($S_{ij}$; see Sec.\ref{Metrics}) on three IBM QPUs: ibm\_lagos, ibm\_brisbane, and ibm\_fez with 7, 127, and 156 qubits, respectively, and IonQ ionq\_forte \cite{Chen_2024}. As a testing protocol, the EfficientSU2 circuit of qiskit is employed \cite{Javadi-Abhari2024}, the QAOA algorithm in the case of ibm\_fez, and random circuits. With only 1000 samples, the $\rho_{zecs}$ of 4 qubits can be recovered. The different $\rho_{zecs}$ are then utilized to create a map of the fidelity and the entanglement entropy for the entire layout of the device. This information is used as a routing technique to select the best chain of 20 qubits for an optimization application. It is shown that this methodology improves the performance of the application compared to the heaviest qiskit transpilation technique.

The paper is organized as follows. Section~\ref{Sec:Methods} provides a description of CS, ZECS, the experimental setup, and the routing application. In Sec.~\ref{Sec:Results}, the results of ZECS on real QPUs and for the routing and non-local correlation applications are presented. Finally, Sec.~\ref{Sec:Conclusions} provides some conclusions.

\section{Methods}\label{Sec:Methods} 
\subsection{Classical Shadow}

Classical shadow is a method for reconstructing an approximate classical description of a quantum system using a small number of measurements \cite{Huang2020}. To reconstruct an $n$-qubit quantum state $\rho$ using N snapshots, random unitary gates $U_i$ are applied to $\rho$

\begin{equation}\label{Eq:UpU}
\rho\rightarrow U_i\rho U_i^\dagger,
\end{equation}
and measured on the computational basis. This results in a bitstrings $|b\rangle \in \{0,1\}^n$ and is modeled by the quantum channel 
\begin{equation}
\mathbb{E}\left[ U_i^\dagger|b_i\rangle \langle b_i|U_i\right] = \mathcal{M}(\rho_i), 
\end{equation}
where the operator $\mathcal{M}$ depends on the set of random unitary transformations $U_i$. A classical snapshot $\rho_i$ of $\rho$ can be constructed using the inverted operator such that
\begin{equation}\label{EQ:Snapshot}
\rho_i = \mathcal{M}^{-1}(U_i^\dagger|b_i\rangle \langle b_i|U_i),
\end{equation}
where $\mathcal{M}^{-1}(X) = (2^n + 1)X - \mathrm{I}$. This is not completely positive, but the collection of $N$ snapshots is expressive enough to predict many properties of the quantum state. Classical shadow is the process of repeating Eq.~(\ref{EQ:Snapshot}) $N$ times, which mathematically is expressed as
\begin{equation}
    S(\rho, N) = \left\{
    \begin{aligned}
        &\rho_1 = \mathcal{M}^{-1}(U_1^\dagger |b_1\rangle \langle b_1| U_1), \\
        &\rho_2 = \mathcal{M}^{-1}(U_2^\dagger |b_2\rangle \langle b_2| U_2),\\
        &\ldots \\
        &\rho_{N} = \mathcal{M}^{-1}(U_{N}^\dagger |b_{N}\rangle \langle b_{N}| U_{N})
    \end{aligned}
    \right\}.
\end{equation}
This method is restricted here to the Pauli basis measurement (see Eq.~(S44) in \cite{Huang2020}) so that each snapshot is given by
\begin{equation}
\rho_i = \otimes_{j=1}^n\left( 3U_j^\dagger |b_j\rangle \langle b_j| U_j - \mathrm{I} \right).
\end{equation}
where $U_j$ changes the basis to the $\{X, Y, \mathrm{or} \ Z\}$ basis. The reconstruction of the state operator $\rho$ using classical shadow is then found from
\begin{equation}\label{Eq:CS}
\rho_\text{cs} = \frac{1}{N}\sum_{i=1}^N \rho_i.
\end{equation}
It is important to note that the inverted channel does not represent a physical system, i.e., it is not a completely positive and trace-preserving (CPTP) channel. However, a sufficiently large Classical Shadow will approximate the true density state operator.

\subsection{ZECS}
In general, $\rho_{cs}$ is a hermitian but not necessarily a positive semi-definite matrix. This means it cannot represent a  quantum state. Additionally, in the current stage of quantum computation, QPUs are inherently noisy, and, therefore, snapshots are partially corrupted by noise. The ZECS methodology is proposed here to mitigate both problems. ZECS has the objective of reconstructing the closest representation of a pure state $\rho$ using the information of $\rho_{cs}$.

Singular value decomposition (SVD) is used to decompose $\rho_\text{cs}$ such that
\begin{equation}\label{Eq:ze1}
    \rho_\text{cs} = \mathrm{V} \Sigma \mathrm{V}^\dagger,
\end{equation}
where  $\Sigma = \mathrm{diag}(|\lambda_1|, |\lambda_2|, ..., |\lambda_n|)$ is a diagonal matrix with the non-negative singular values  of $\rho_{cs}$  in decreasing order, These coincide with the absolute values of the eigenvalues because $\rho_{cs}$ is hermitian. $V = [|\psi_1\rangle, |\psi_2\rangle, ..., |\psi_n\rangle]$ is a matrix with the (orthonormal) eigenvectors of $\rho_\text{cs}$. The zero-entropy (ZE) step consists of truncating the information used to reconstruct the density state operator to only that of the largest eigenvalue. This step is analogous to the rank reduction in QST \cite{gross2010, Baldwin_2016, franca2021} and can be seen as completely removing the entropy of $\rho_{cs}$. The eigenvector $|\psi_1\rangle$ associated with the largest singular value of $\rho_{cs}$, i.e., $|\lambda_1|$ is then used to reconstruct the density state operator by
\begin{equation}\label{Eq:ze2}
    \rho_\text{zecs} = |\psi_1\rangle \langle\psi_1|.
\end{equation}
This new density state operator fulfills the positive semidefinite and unit-trace conditions for representing the approximate density state operator $\rho$. 

Note that by the well-known Mirsky generalization of the Eckart-Young theorem \cite{Golub1987}, the closest  approximation of $\rho_{cs} $ by means of a rank-one hermitian operator $A$ with respect to a unitarily invariant operator norm is given by $A_1=\mathrm{V} \Sigma_1 \mathrm{V}^\dagger=|\lambda_1|\,\rho_\text{zecs}$ where $\Sigma_1 = \mathrm{diag}(|\lambda_1|, 0, ..., 0)$, i.e.,
\begin{equation}\label{Eq:ey1}
    \|\rho_\text{zecs}\,|\lambda_1|-\rho_{cs}\| = \inf_{ \text{rank}(A)=1}  \|A-\rho_{cs}\|.
\end{equation}
In general, $A_1$ is non-negative but not unit trace. Therefore, to obtain a density operator, $A_1$ is renormalized so that $\rho_\text{zecs}=A_1/\Tr(A_1)$. The renormalization step modifies the operator to meet the trace condition but compromises the norm minimization. However, the loss with respect to Eckart-Young optimality vanishes in the limit as $\rho_\text{cs}$ is not too far from purity, i.e., if $1-\lambda \ll 1$, which is fulfilled in the experiments we analyze here. In return for this slight compromise, $\rho_\text{zecs}$ provides the best representation of $\rho_\text{cs}$, which fulfills the \textit{purity} characteristic of the theoretical output of the quantum circuit.

Numerical evidence is provided in Sec.\ref{Metrics} that shows the zero entropy strategy to recover information of a mixed state. We use a random perturbation on a 2-qubit Bell state following the methodology in 
\cite{Montanez-Barrera2022} and show that if the perturbation is not large, one can always recover more information using this zero-entropy (ZE) methodology. Evidence is given in terms of the fidelity, F, the trace distance, D, and the concurrence, C, for a description of these metrics.

\subsection{Crosstalk characterization}
Among the various techniques to quantify crosstalk, Simultaneous Randomized Benchmarking (SRB) \cite{Gambetta_2012,mckay2020correlated} is used to assess whether running randomized tests in parallel reduces the performance of otherwise independent systems. SRB consists of three experiments: two involve performing RB on one system while the other remains idle, and the third runs RB on both systems simultaneously. This yields two error rates $r_i$ for the idle case and $r_s$ for the simultaneous case. The crosstalk is then quantified as $r_i-r_s$, representing the change in RB performance of each subsystem due to concurrent operation.

Another way to quantify crosstalk is through gate set tomography (GST), which can be used to determine whether the observed data is better explained by a crosstalk-free model, a model with context-dependent crosstalk between nearby qubits, general crosstalk, or non-Markovian noise. The most accurate representation is identified based on a metric called the likelihood \cite{Rudinger_2021}.

In this work, we use the entanglement entropy to characterize crosstalk.
The entanglement entropy is expressed as 
\begin{equation}\label{Eq:Sab}
    S_{ab} = -\rho_a \log(\rho_a) = -\rho_b \log(\rho_b),
\end{equation}
where $\rho_a=\Tr_b(\rho)$ is the partial trace of the composite density state operator $\rho$ over the basis states of subsystem $b$. The classical cost to compute the entanglement entropy from a given state $\rho$ is dominated by two steps. The first entails forming the reduced density matrix $\rho_{a}=\operatorname{Tr}_{b}\rho$, which requires $O(d_a^2 d_b)$ arithmetic operations, where $d_a=2^{n_q^a}$ and $d_b=2^{n_q^b}$ are the Hilbert-space dimensions of subsystems $a$ and $b$. The second consists of obtaining the spectrum required by Eq.~\ref{Eq:Sab} (e.g., via SVD or diagonalization of $\rho_a$), which costs $O(d_a^3)$. 

When subsystems $a$ and $b$ are disentangled, i.e., $\rho = \rho_a \otimes \rho_b$, $S_{ab} = 0$. In our experiments, we run quantum circuits on subsets of qubits independently, reconstructing $\rho$ using ZECS, so the entanglement entropy  of combinations of these subsystems is expected to be zero. Uncorrelated noise should not affect the entanglement entropy. However, if crosstalk induces correlations between the subsystems, the entanglement entropy will deviate from zero

\begin{figure}[!t]
\centering
\includegraphics[width=8.5cm]{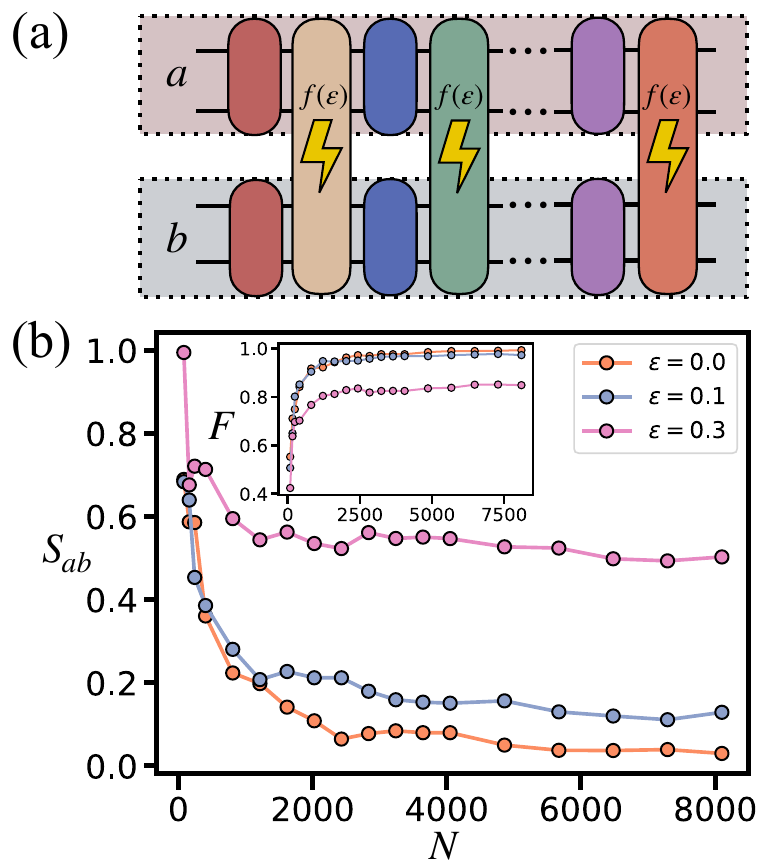}
\caption{\label{Fig:crosstalk} Crosstalk simulation between subsystems $a$ and $b$. (a) A random circuit used to simulate crosstalk between subsystems $a$ and $b$. The noisy gate that represents the crosstalk connects both subsystems and has a strength parameter $\varepsilon$. (b) Entanglement entropy versus the number of shadows used for different crosstalk strengths for the simulation of the circuit in (a). The inset plot shows the fidelity for the same process.} 
\end{figure}

Figure~\ref{Fig:crosstalk} shows how ZECS reconstruction can be used to identify crosstalk in noisy quantum circuits. Fig.~\ref{Fig:crosstalk}(a) depicts the circuit composed of two subsystems, $a$ and $b$. Crosstalk is simulated by introducing additional random two-qubit gates acting between one qubit of subsystem $a$ and one qubit of subsystem $b$, with the interaction strength controlled by the parameter $\varepsilon$. We consider random circuits with a two-qubit depth of 10, simulated using Qiskit’s noisy simulator \cite{ibm_build_noise_models} with 2-qubit depolarizing noise $\lambda=1\times 10^{-3}$. At each depth step, a random two-qubit gate is applied to each subsystem, followed by a crosstalk gate. Three scenarios are investigated: $\varepsilon=0$ (no crosstalk), $\varepsilon=0.1$, and $\varepsilon=0.3$. The crosstalk gates are implemented as $R_{xx}(\theta)$ and $R_{zz}(\theta)$ rotations, with $\theta$ randomly chosen in the interval $[-\varepsilon, \varepsilon]$. As $\varepsilon$ increases, the effective strength of the induced crosstalk also increases.

Figure~\ref{Fig:crosstalk}(b) shows how the entanglement entropy changes with the number of shadows for three different crosstalk strengths. Since depolarizing noise is included in all cases, the entanglement entropy approaches zero when $\varepsilon = 0$, indicating that uncorrelated noise does not produce a large $S_{ab}$. In contrast, increasing the crosstalk strength consistently increases the observed $S_{ab}$, showing that ZECS is an effective method for detecting crosstalk. The inset of Fig.~\ref{Fig:crosstalk}(b) presents the corresponding fidelity results, where stronger crosstalk also leads to lower fidelity. Together, these simulation outcomes clarify why ZECS serves as a useful methodology for identifying crosstalk.

\subsection{Experiments}

To evaluate the ZECS methodology on real quantum hardware, we use the IonQ Forte (ionq\_forte) and the IBM Falcon (ibm\_lagos), Eagle (ibm\_brisbane), and Heron (ibm\_fez) processors \cite{IBMQuantumProcessors, IBM2021}. ionq\_forte is a trapped ion device with a fully connected layout, while the other QPUs are superconductive quantum processors \cite{Koch2007} with a Heavy-Hex layout \cite{IBMHeavyHexLattice}. 

\begin{figure}[!t]
\centering
\includegraphics[width=8.5cm]{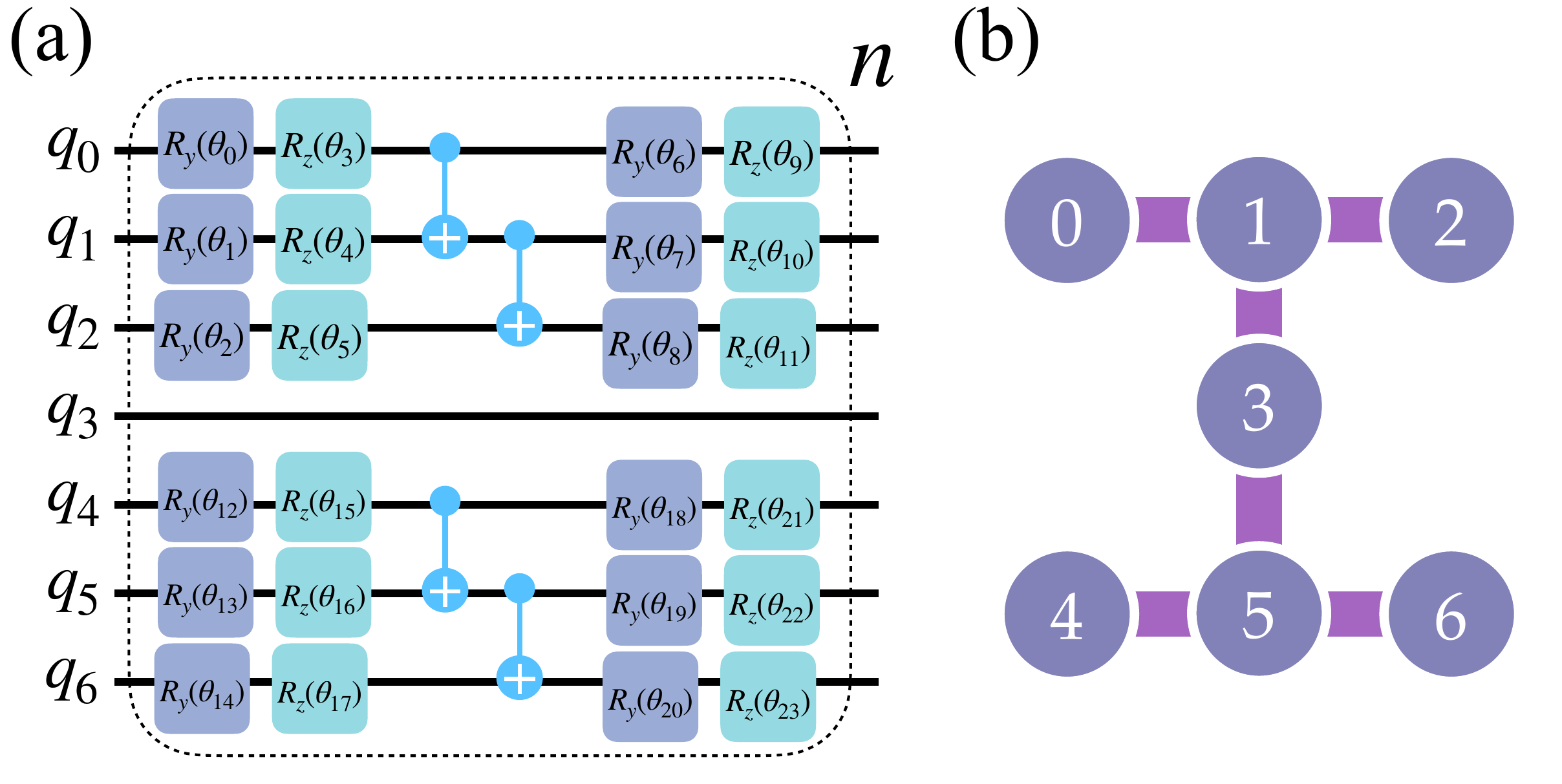}
\caption{\label{Fig:circuit_used} (a) The EfficientSU2 circuit used for the characterization of ibm\_lagos. The circuit is repeated $n$ layers, and the parameters for each layer, the $\theta_i^n$, are randomly chosen from a uniform distribution between $0$ and $\pi/2$; (b) The layout of {\it ibm\_lagos} where the vertices represent the qubits and the edges the physical connectivity between them.}
\end{figure}

Fig.~\ref{Fig:circuit_used} shows (a) the circuit used for the ibm\_lagos experiment and (b) the ibm\_lagos layout. The circuit used is an EfficientSU2 gate from qiskit. This is a parametrized circuit with $4nN_q$ parameters, where $n$ is the number of repetitions of the circuit and $N_q$ is the number of qubits. The parameters are randomly selected from a uniform distribution with values between 0 and $\pi/2$. 10,000 snapshots are collected for this circuit on ibm\_lagos, measuring the 7 qubits involved.

In the case of ibm\_brisbane, 1 and 10 repetitions of the EfficientSU2 gate are applied in pairs of qubits throughout the device. The parameters are randomly selected for different pairs of qubits using the same distribution employed with ibm\_lagos. 6,000 snapshots are utilized to measure all the qubits at the end of the protocol. In the case of ibm\_fez and ionq\_forte, the 2-qubit systems execute an LR-QAOA \cite{montanezbarrera2024universal} protocol with p=10. This circuit consists of 20 CZ gates and 10 Rx gates.

\subsection{Routing application}\label{routing}

The entanglement entropy, $S_{ij}$, of non-local pairs of qubits can be used to detect leakage of information in a non-local way. It is believed that this non-local crosstalk in at least some cases can occur at the multiplexing readout stage \cite{Jurcevic2021, Jerger2012}. To determine if this information gained through ZECS provides a better strategy for routing qubits in the IBM Heavy-Hex topology than the method used by the qiskit transpiler, the LR-QAOA is used. LR-QAOA is employed in combinatorial optimization applications and is an approximation of a digital adiabatic protocol consisting of a fixed set of linear annealing parameters in the QAOA \cite{Farhi2014, Willsch2022, montanezbarrera2024universal, hess2024effectiveembeddingintegerlinear, Kremenetski2023}.

For routing, qiskit uses an algorithm called VF2Postlayout \cite{Nation_2023} that consists of finding the best isomorphic subgraphs to the input circuit using a heuristic objective function derived from the calibration data of the device. In qiskit, there are four levels of routing optimization (from 0 to 3), with 3 being the level where the most effort in the algorithm occurs to find the layout with the lowest error. 

To compare both methods, a weighted maxcut problem (WMC) is used with weights randomly selected from the options [0.1,0.2,0.5,2.0] on a 20-qubit 1D-chain topology. LR-QAOA is run from $p=3$ to 100 layers with a $\Delta_{\gamma, \beta} = 1.0$ (see Eq. (4) in \cite{montanezbarrera2024universal}). The ZECS and qiskit methodologies are run consecutively on ibm\_brisbane with 1,000 shots. The performance is calculated using the approximation ratio (see Eq. (11) in \cite{montanezbarrera2024universal}).

\section{Results}\label{Sec:Results}
Results for the experiments conducted on the QPUs are presented below.

\subsection{ibm\_lagos}

Fig. \ref{Fig:lagos} shows a comparison of the infidelity versus the number of snapshots resulting from the CS and ZECS reconstruction of the density state operator for a 3-qubit EfficientSU2 protocol with 7 repetitions. Two different samplers are used: a noiseless sampler (qasm\_simulator) and ibm\_lagos to show the effect of noise on the reconstruction. Under the assumption of a noiseless procedure (green lines), at $N=1,000$ snapshots, the qasm\_simulator reaches an infidelity of $\approx 0.001$ for both CS and ZECS. In the case of a real device (orange curves), using CS, the infidelity stabilizes at around $0.25$ while using ZECS, more of the noise is removed, obtaining an infidelity of $0.057$.  An extended study of the noise on ibm\_lagos is presented in Appendix \ref{A:ibmlagos}.

\begin{figure}
\centering
\includegraphics[width=8.5cm]{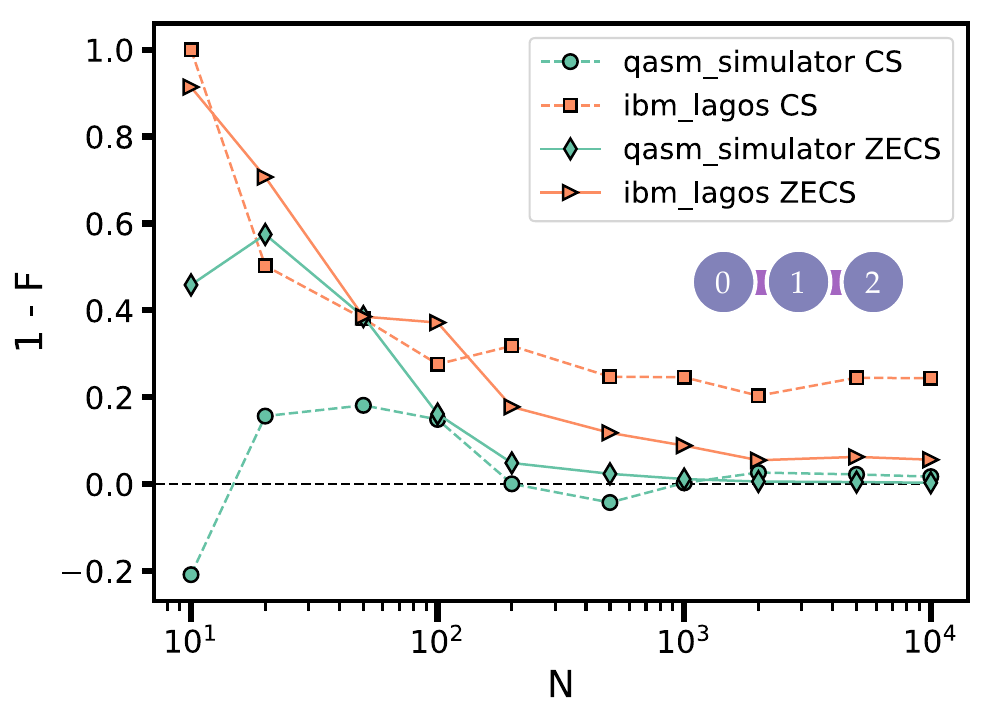}
\caption{\label{Fig:lagos} Infidelity versus the number of snapshots using the qasm\_simulator and ibm\_lagos for 7 repetitions of the protocol of Fig. \ref{Fig:circuit_used}(a) on qubits 0, 1, and 2. The dashed lines represent the CS and the solid lines the ZECS reconstructions of the density state operator.}
\end{figure}

\subsection{ibm\_brisbane}
\begin{figure}[!tbh]
\centering
\includegraphics[width=8.5cm]{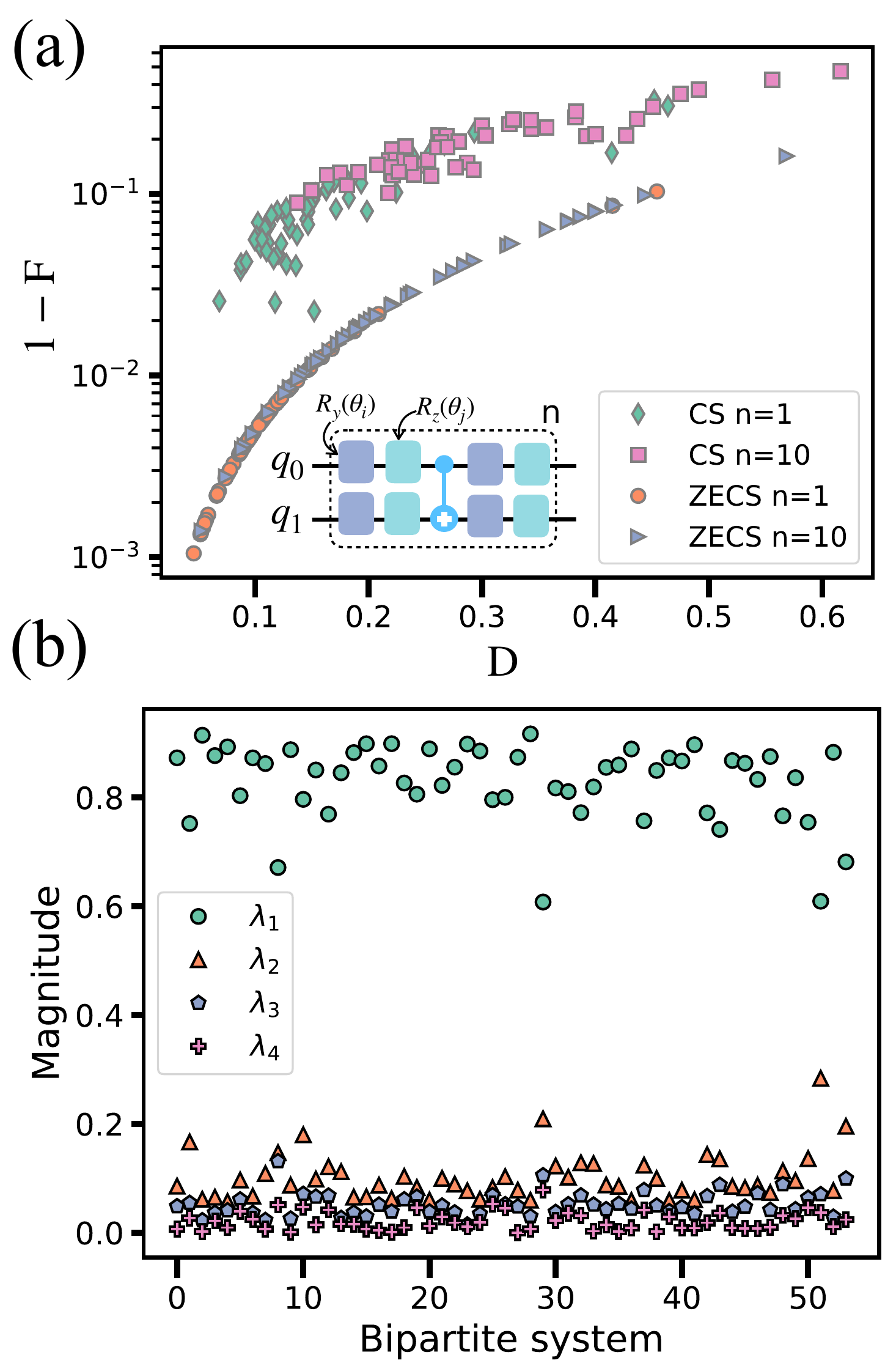}
\caption{\label{Fig:brisbane} (a) ($1-F$) vs. $D$ using CS and ZECS for the 2-qubit experiment of the EfficientSU2 gates for $n=1$ and 10 repetitions on ibm\_brisbane. The inset circuit is the EfficientSU2 gate repeated n times with random parameters $\theta_i$. Every marker represents a pair of qubit results. (b) Singular values of $\rho_{cs}$ for the bipartite subsystems in the $n=10$ EfficientSU2 experiment.}
\end{figure}

Figure ~\ref{Fig:brisbane}(a) shows the relation between the infidelity, $1 - F$, and the trace distance, $D$ (See Sec.\ref{Metrics}), for two experiments, 1 and 10 layers of EfficientSU2 in pairs of qubits of {\it ibm\_brisbane} using CS and ZECS. In Table \ref{tab:qubit_cases}, there is a list of the qubits involved in the EfficientSU2 experiments. In both metrics, a value of zero is desired, and the ZECS can correct most of the noise, bringing the error to an isentropic line where $1 - F$ and $|D|$ are correlated. In contrast, in the CS case, because of an unstable reconstruction of the density state operator, there is no strong correlation between the two metrics. Figure ~\ref{Fig:brisbane}(b) shows the singular values magnitude for the bipartite systems involved in the $n=10$ EfficientSU2 experiments. The order of them is the same as shown in Table \ref{tab:qubit_cases}. Note that there is a good separation of $\lambda_1$ to the other singular values. This is a good indication that noise is not strong enough to forbid a good reconstruction of the density state operator.

\begin{figure*}[t]
\centering
\includegraphics[width=18cm]{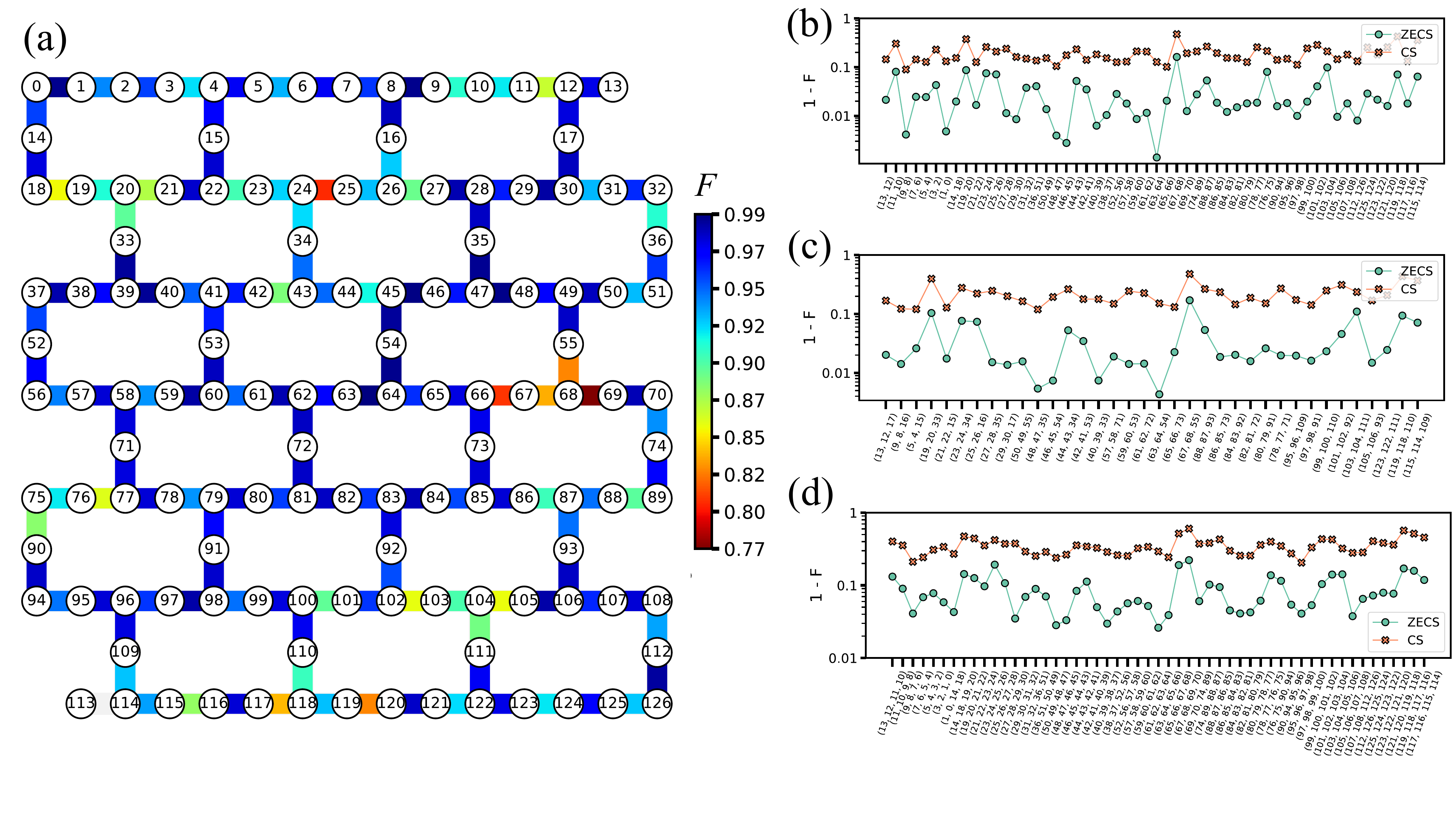}
\caption{\label{Fig:brisbane_layout} Fidelity information of ibm\_brisbane based on the reconstruction of the density state operator using CS and ZECS. (a) Fidelity map of the different 2, 3, and 4-qubit subsystems. In the case of 3-qubit subsystems, the fidelity is represented by the edge between the idle qubit node and its neighbor. For example, in (13,12,17), the edge between nodes 12 and 17 represents the fidelity. In the case of the 4-qubit subsystem, the edge that connects the two pairs of qubits represents the fidelity, e.g., in (13,12) and (11,10), the edge between nodes 12 and 11 represents the fidelity. (b)-(d) 1-F versus the subsystems involved in the reconstruction of the density state operator using ZECS and CS. (b) Qubits are involved in the EfficientSU2 protocol. (c) The 3-qubit subsystems involve an idle qubit. For example, in subsystem (13,12,17), qubit 17 is the idle qubit. (d) 2 neighbor pairs of 2-qubit subsystems that are involved in the EfficientSU2 protocol.}
\end{figure*}

Figure ~\ref{Fig:brisbane_layout} shows the fidelity information of {\it ibm\_brisbane} obtained from the reconstruction of density state operators using $N=6,000$ snapshots and the ZECS methodology for 10 repetitions of the EfficientSU2 circuit. Fig \ref{Fig:brisbane_layout}(a) shows the layout of {\it ibm\_brisbane}, with edge color representing the fidelity of the density state operator reconstructed. This reconstruction is based on the information of 2, 3, and 4-qubit density state operators. In Table \ref{tab:qubit_cases}, there is a list of the qubits involved in each reconstruction. This fidelity map helps to identify regions where poor performance is present. For example, qubit 68 and its neighboring nodes have the lowest performance. If we compare this fidelity against, for example, the RB fidelity, we can expect a more detailed noise characterization. Because the characterization of 2-qubit gates using RB uses only one expectation value. 

Fig.\ref{Fig:brisbane_layout}(b) shows the infidelity of the 2-qubit systems involved in the EfficientSU2 protocol. There is a significant impact of applying the ZECS methodology to the CS information, reducing the infidelity from 0.19($\pm 0.08$) to 0.03($\pm 0.03$) on average. Furthermore, there is a correlation between the infidelity of CS and ZECS, but not in all cases. For instance, subsystem (27,28) has a lower infidelity than subsystem (26,25) using the CS reconstruction, but once ZECS is applied, subsystem (26,25) has a lower infidelity. This suggests that some kinds of errors are more prone to be corrected. ZECS shows improvements in some cases of 1 order of magnitude in the infidelity compared to CS. It is interesting to note that ZECS is based completely on the CS information and no further assumptions are made. Therefore, the extra information recovered in ZECS comes completely from the QPU. Figs. \ref{Fig:brisbane_layout}(c) and (d) show the infidelity for the 3-qubit and 4-qubit cases. The recovery of ZECS is also important in these cases. A detriment in fidelity from the 3 and 4-qubit cases when compared with the 2-qubit case results primarily from the dimensionality of the density matrix that is reconstructed. In the case of the 2-qubit problem, there are only $2^{N_q} = 4$ complex parameters, while for 3 qubits, there are 8 and 4 qubits 16.

\begin{figure}
\centering
\includegraphics[width=8.5cm]{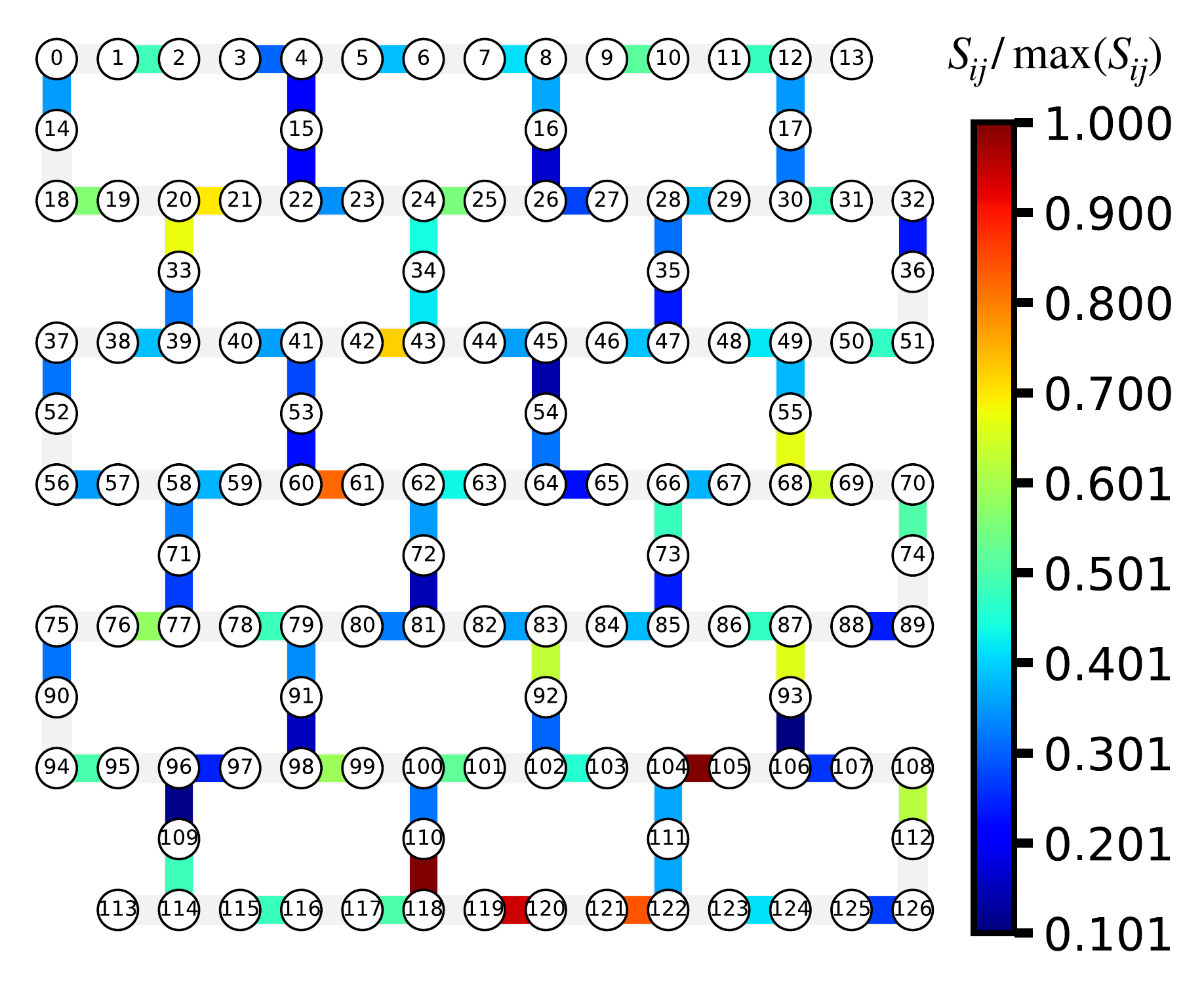}
\caption{\label{Fig:ent_entropy} Entanglement entropy between subsystems not connected during the CS experiment. The color map represents the normalized entropy of the system. The entropy is normalized with respect to the maximum entropy of the 3-qubit and the 4-qubit systems. A red color is an indication of an unwanted correlation between subsystems. For instance, the edge connecting qubits 12 and 17 yields $S_{12,13} = S_{17}$ where the two subsystems are (12,13) and (17). Because these subsystems are not connected in the protocol, their entanglement entropy should be zero so if any entropy is found, it could be related to the presence of some crosstalk. }
\end{figure}

Figure \ref{Fig:ent_entropy} shows the entanglement entropy, $S_{ij}$, of the different subsystems. Edges are associated with 3-qubit and 4-qubit subsystems. In Fig. \ref{Fig:brisbane_layout}(a), fidelity is presented as a quantifier of the quality of the 2-qubit operations and the 3-qubit and 4-qubit interactions. However, correlating this information with unwanted crosstalk is limited. In contrast, the entanglement entropy, $S_{ij}$, between the subsystems indicates that there is a leak of information between the subsystems. In the ideal case, $S_{ij}=0$ because the subsystems are not interacting. Furthermore, information about the fidelity and the entanglement entropy seems to be correlated, i.e., in regions where there is low fidelity, the subsystems have high $S_{ij}$. But the region with the worst fidelity (68,69) is not the region with the highest $S_{ij}$. In the lower part of ibm\_brisbane, there is higher entanglement entropy in qubits 104, 105, 110, 118, 121, and 122. 

\begin{figure*}
\centering
\includegraphics[width=18cm]{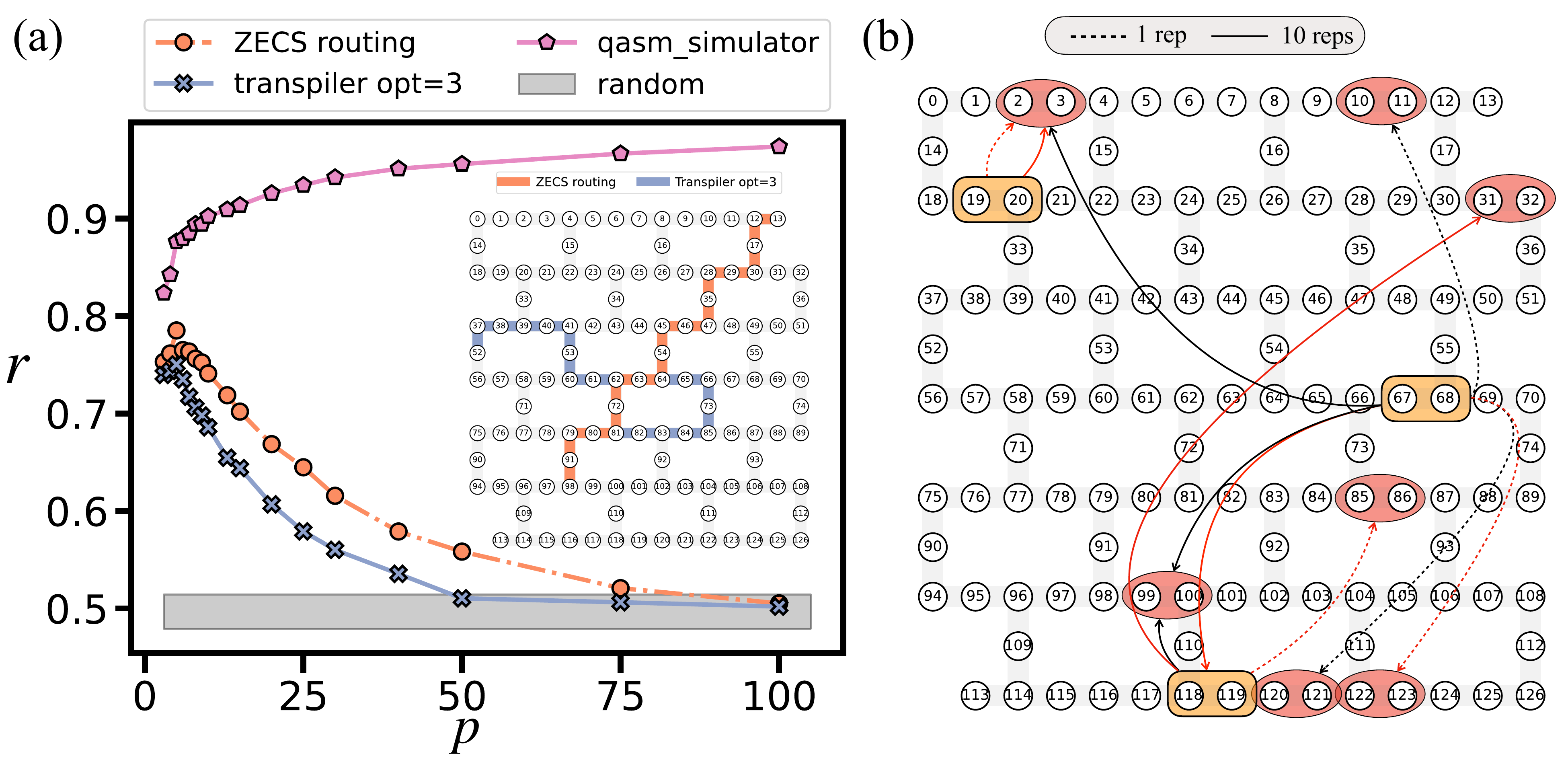}
\caption{\label{Fig:qaoa} Combinatorial optimization application using the qubit chain found via ZECS versus that determined via the best transpilation method of qiskit. (a) Approximation ratio versus the number of LR-QAOA layers $p$ of a random WMC using both chains. In a noiseless simulation, i.e., qasm\_simulator, as $p$ grows, the solution approaches $r=1$. The shaded region, random, shows where the outputs from the QPU cannot be distinguished from a random sampler. The inset shows the chain of qubits used in the application for the two methods. There is an overlap of qubits 62, 63, and 64 for both routing methods.  (b) Non-local map of high $S_{i,j}$ for three subsystems of interest: (19,20), (67,68), and (118, 119). The orange squares represent the subsystems of interest. The pink ovals represent the subsystems that share a high entanglement with the subsystems of interest. Dashed and solid lines are the results from 1 and 10 repetitions of the EfficientSU2 protocol, respectively. The arrows connect the systems that have an $S_{i,j}$ 2 standard deviations above the mean value over all possible pairs. The red arrows represent the highest $S_{i,j}$ for each subsystem analyzed.}
\end{figure*}

Now, based on the information provided by Figs. \ref{Fig:brisbane_layout} and \ref{Fig:ent_entropy}, the best 20 qubits that form a 1D chain are chosen and compared to the best routing procedure of Qiskit. In the inset of Fig. \ref{Fig:qaoa}(a), the chain chosen using ZECS information is indicated in orange, while the chain chosen by the qiskit transpiler with an optimization level 3 is shown in blue. The routing procedure of qiskit uses calibration information of the device, including the native 2-qubit gates error. In the case of ZECS, the fidelity and entanglement entropy information are used to guide the routing. 

Table \ref{tab:1} summarizes the mean values for 2-qubit error of the native ECR gate of ibm\_brisbane ($\mathrm{ECR}_{error}^{avg}$), the average fidelity $\bar{F}$, and the average entanglement entropy $\bar{S_{ij}}$ of the qubits selected using both routing methodologies. In parentheses is the standard deviation. In terms of $\mathrm{ECR}_{error}^{avg}$, the error of the qubits chosen by qiskit is much lower than the ZECS error. In terms of $\bar{F}$ and $\bar{S_{ij}}$, the ZECS routing is better than the qiskit routing.

\begin{table}[h]
    \centering
    \caption{Comparison of the ZECS and Qiskit Transpilation Routing.}
    \label{tab:1}
    \begin{tabular}{@{}lll@{}}
        \hline
        \textbf{Metric} & \textbf{ZECS Routing} & \textbf{Qiskit Routing} \\ 
        \hline
        $\mathrm{ECR}^{avg}_{error}$ & $0.01 \ (\pm 4.8\times 10^{-3})$ & $0.006 \ (\pm 1.8\times 10^{-3})$ \\ 
        $\bar{F}$ & $0.983 \ (\pm 0.011)$ & $0.974 \ (\pm 0.015)$ \\ 
        $\bar{S_{ij}}$ & $0.299 \ (\pm 0.09)$ & $0.375 \ (\pm 0.155)$ \\
        \hline
    \end{tabular}
\end{table}

Figure \ref{Fig:qaoa}(a) shows the results of the optimization solutions using the LR-QAOA of Sec. \ref{routing} for the WMC. The y-axis represents the approximation ratio vs. the number of QAOA layers. This metric indicates the proximity of the QPU's outputs to the optimal solution of the problem, with $r=1$ indicating a 100\% probability of finding the optimal solution of WMC using LR-QAOA. In an ideal case, as $p$ grows, $r$ approaches 1. However, when noise is involved, $r$ grows to the point where the LR-QAOA is stronger than the noise inherent in the device. As can be seen, using the ZECS routing, the chosen set of qubits improves the lifetime of the LR-QAOA compared to the routing chosen by qiskit by more than 33\%, i.e., from $p=50$ to $p=75$, and the peak of the approximation ratio from $r=0.750$ to $r=0.785$, which is a 10\% reduction in error when compared to the ideal approximation ratio of $r=0.876$ at $p=5$. This suggests that the qiskit routing and the two-qubit errors do not fully capture the characteristics needed for a set of qubits to perform well on some tasks that require cohesion between qubits. 

Fig. \ref{Fig:qaoa}(b) shows another application of ZECS, i.e., the detection of non-local correlations in pairs of qubits. Because of their low performance, three subsystems of interest  are analyzed: (19,20), (67,68), and (118,119). In this case, the density state operators involving the 2-qubit subsystems of interest are reconstructed, as are the other subsystems used in the EfficientSU2 protocol for 1 and 10 repetitions that do not share a direct connection. The non-local interactions could show potential crosstalk coming from, for example, the multiplexing readout. The strongest entanglement entropy, $S_{i,j} = 0.237$, is between pairs (19,20) and (2,3), which is 3.5 $\sigma$ above the mean entropy of $\bar{S}_{i,j}=0.113(\pm 0.035)$ of all pairs versus (19,20) for the 10 repetitions case. Note that the $F$ of (2,3) (Fig.\ref{Fig:brisbane_layout}(a)) is not affected, which suggests that the information of (2,3) corrupts that of (19,20) but not vice versa. The case of (67, 68) qubits is the most correlated subsystem, especially with the bottom region of the QPU, i.e., subsystems (118, 119), (120,121), and (122,123). The detriment in $F$ of the subsystems of interest might be related to this unwanted non-local interaction.

\subsection{ibm\_fez}
\begin{figure*}
\centering
\includegraphics[width=17.5cm]{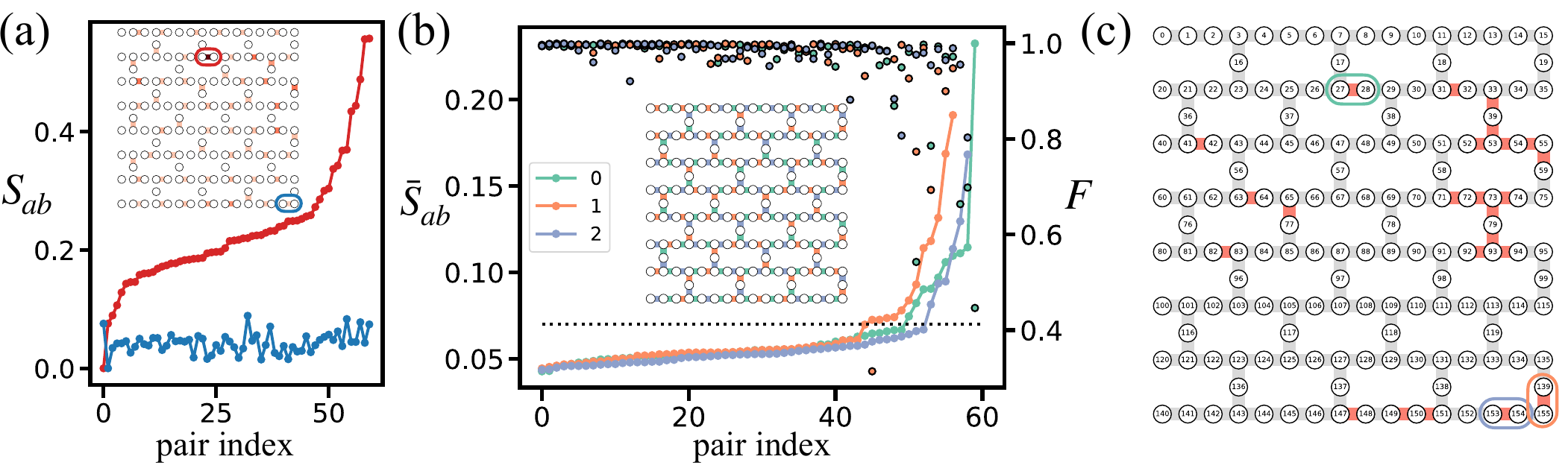}
\caption{\label{Fig:ibm_fez} Entanglement entropy of pairs of qubits on ibm\_fez. (a) The red line shows the pair of qubits that shares the largest $\bar{S}_{ab}$ with the other pairs, and the blue line shows the pair of qubits with the lowest $\bar{S}_{ab}$ in experiment 0. The inset graph shows the pairs of qubits in experiment 0, with edges describing the $\bar{S}_{ab}$ of each pair, and a reddish color indicating a stronger $\bar{S}_{ab}$. The two ovals represent the two pairs of qubits with minimum and maximum correlation. (b) $\bar{S}_{ab}$ (solid line) for the three experiments: the inset layout shows the distributions of pairs of qubits in each experiment. The circles represent the fidelity (right y-axis) of the pair of qubits experiment. The dotted line represents the $\bar{S}_{ab}$ that is considered normal. (c) ibm\_fez layout with orange lines representing the $\bar{S}_{ab}$ that are above the dotted line in (b) and, therefore, the pairs of qubits that generate a high entanglement when used.}
\end{figure*}

To systematically explore the scalability of non-desired entanglement in larger quantum systems, the \texttt{ibm\_fez} quantum processor is employed using three experimental configurations consisting of three layers of qubit pairs (see device layout in Fig.~\ref{Fig:ibm_fez}(b)).In each configuration, 6,000 measurement shots are performed to reconstruct 4-qubit subsystems composed of two independent qubit pairs. Since the LR-QAOA circuit with p=10 operates separately on each pair, the entanglement entropy within a single pair should ideally be zero. A nonzero entropy value indicates residual correlations between the subsystems, suggesting the presence of crosstalk or other unintended interactions.

Figure~\ref{Fig:ibm_fez}(a) presents the results for the 2 cases: in red (blue) the pair with the highest (lowest) entanglement entropy relative to the rest of the systems. While the pair in blue exhibits near-zero entanglement entropy with the other subsystems, the red pair reaches a high entanglement entropy with all other subsystems. Notably, when a qubit pair shows significant entanglement, it is not restricted to interactions with a few other pairs but instead displays a more global entanglement across the full QPU. This widespread entanglement can be detrimental to the QPU performance.

Figure~\ref{Fig:ibm_fez}(b) presents the average entanglement entropy (left) and fidelity (right) for the qubit pairs involved in the three experimental configurations. The edge color coding in the inset corresponds to the respective experimental setups. In all three cases, a small subset of qubit pairs exhibits significantly higher entanglement compared to the others. Interestingly, while low fidelity sometimes correlates with high entanglement entropy, this is not a consistent trend. Additionally, it is observed that highly entangled pairs can also exhibit high fidelity. This indicates that entanglement cannot be reliably predicted solely based on two-qubit gate error rates. Figure \ref{Fig:ibm_fez} shows in orange the systems with a high $\bar{S}_{ab}$. This gives a map of the qubits that must be avoided in experiments. The ovals represent the systems with the highest $\bar{S}_{ab}$ for each experiment.

\subsection{Cross-platform comparison}

\begin{figure}
        \centering
        \includegraphics[width=1\linewidth]{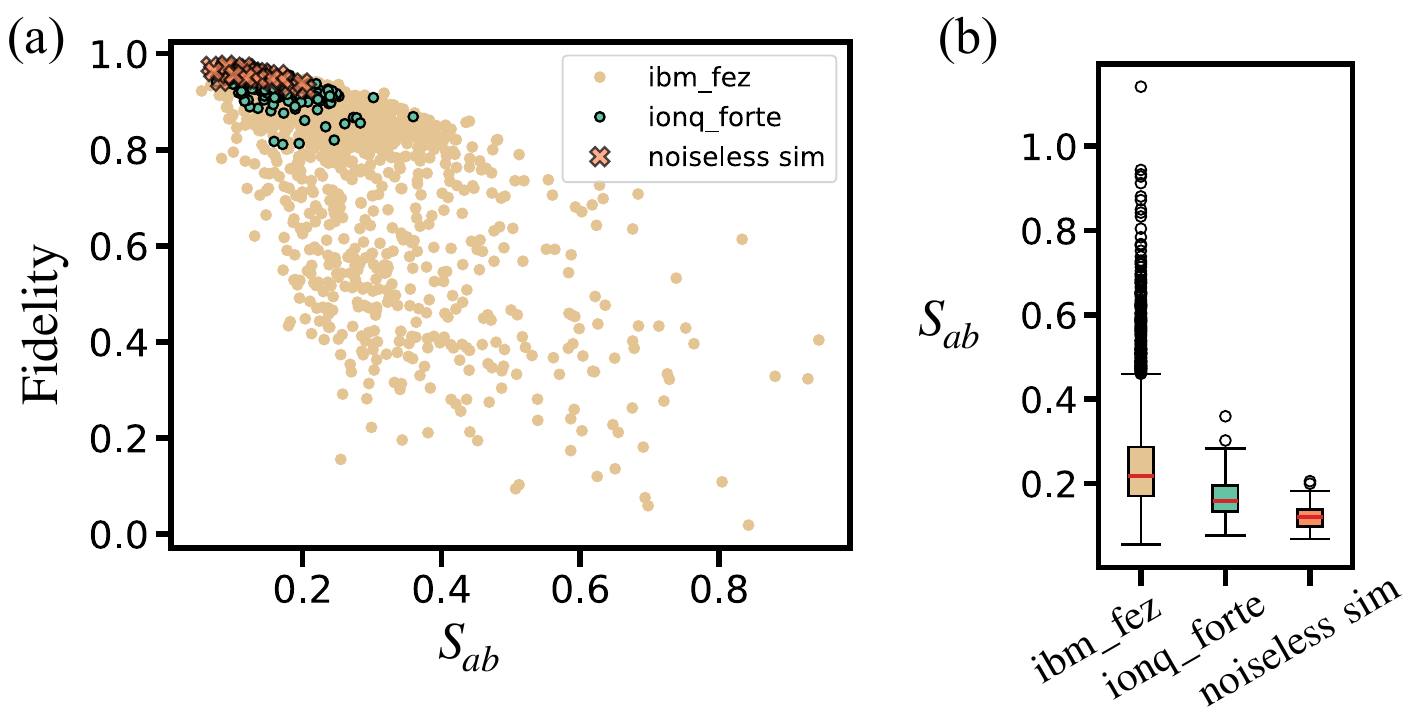}
        \caption{Cross-platform comparison of ZECS between trapped ions and superconducting QPUs of 2-qubit circuit using 1000 samples. (a) Fidelity versus entanglement entropy. Markers represent all the 2-subsystem pairs of each device. (b) The box shows the interquartile range (IQR) for the entanglement entropy, with the central red line indicating the median. Whiskers extend to data within 1.5IQR, caps mark their endpoints, and circles are outliers.}
        \label{fig:QPU-comparison}
    \end{figure}

Figure~\ref{fig:QPU-comparison} shows the comparison between ionq\_forte, ibm\_fez, and a noiseless simulator. From the previous experiment on ibm\_fez, result reconstructions with only 1000 samples are used in 60 pairs out of 176 possible pairs, and the same protocol is run on ionq\_forte using its 36 qubits (18 pairs) and on the noiseless simulator using 20 qubits (10 pairs). The reconstruction is then made in pairs of 2-qubit subsystems throughout all possible pairs, which grows as $N_{total} = N_{pairs}(N_{pairs}-1)/2$. In the case of ibm\_fez, there are 1770 pairs, on ionq\_forte 153 pairs, and on the noiseless simulator 45 pairs. 

Figure~\ref{fig:QPU-comparison}(a) presents the entanglement entropy versus fidelity for the three scenarios. Among them, ibm\_fez shows the strongest impact of crosstalk, affecting many qubit pairs. Note that some part of the entanglement entropy can still be attributed to sampling limitations since even in the noiseless simulation, a residual $S_{ab}$ remains.

Figure~\ref{fig:QPU-comparison}(b) summarize the $S_{ab}$ of the different platforms. The median $S_{ab}$ values (with the upper whiskers in parentheses) are $[0.22(0.46), 0.16 (0.28), 0.12(0.18)]$ for the ibm\_fez, ionq\_forte, and noiseless case, respectively. Furthermore, ibm\_fez shows 190 outliers, compared with only 2 outliers in the other two cases. These findings are consistent with previously reported performance across both platforms when running the same quantum applications \cite{montanezbarrera2024universal, montanezbarrera2025evaluating}. Despite ibm\_fez having a lower average two-qubit gate error rate ($2.6\times10^{-3}$) than ionq\_forte ($10.2\times10^{-3}$), the performance of the ionq\_forte in these applications is better.

\section{Conclusions}\label{Sec:Conclusions}

In this work, ZECS has been proposed as a tool to diagnose quantum devices in terms of their operational quality and the leakage of information. ZECS is a methodology used to reconstruct the closest representation of the true density state operator of a quantum circuit, given the information of CS. Similar to rank 1 QST, ZECS considers only the eigenvector associated with the largest eigenvalue of the CS density state operator reconstruction. In this step, the entropy of the circuit is artificially suppressed. This methodology ensures that the density state operator represents a positive semidefinite and unit trace matrix. These conditions are needed to represent a true quantum state. As the number of required samples is expected to grow exponentially, as is the case with other QST methodologies, the focus here is on using ZECS to reconstruct small sections of QPUs for performance evaluation and information leakage detection.

Experimental validation of the ZECS methodology is performed using ibm\_lagos, ibm\_brisbane, ibm\_fez, and ionq\_forte. 1,000 shots are demonstrated to be sufficient to recover the information of 2, 3, and 4 qubit density state operators. Using this information, a diagnostic in terms of fidelity, trace distance, and entanglement entropy is given. These metrics show how pairs of qubits perform when they are involved in gate operations. Entanglement entropy provides additional information about the crosstalk between pairs of qubits. This is, to our knowledge, the first cross-platform quantification of crosstalk under identical circuit conditions.

Using ZECS, regions where quantum devices have low performance are identified and can, thus, be avoided. This methodology is shown to have better performance in finding a set of qubits for a quantum optimization application compared to the qiskit routing procedure, improving the lifetime of the quantum algorithm by more than 33\% and the approximation ratio from $r=0.750$ to $r=0.785$. The improvement is the result of a more informed decision. 

While qiskit routes are based on the collected calibration information of disjoint terms, routing using ZECS uses detailed information from the density state operator, like fidelity and entanglement entropy. These are meaningful metrics that can be correlated to noise and crosstalk. This reconstruction is flexible, i.e., the qubits for the reconstruction can be chosen arbitrarily, although it is necessary to be aware that the number of snapshots needed grows exponentially with the number of qubits involved in the reconstruction. However, with only 6000 shots, a good estimation of the density state operator for up to 4 qubits can be given. 

We extended the ZECS experiments to ionq\_forte, a trapped-ion QPU from IonQ. On this platform, crosstalk is present but less pronounced than on ibm\_fez. The results also reveal that certain subsystems of ibm\_fez are highly correlated, consistent with the large number of outliers observed. This protocol can also be extended to other platforms, such as neutral-atom and photonic-based QPUs. The only requirement is the availability of single-qubit gates that include at least $R_X$ and $R_Z$ rotations.

The density state operator reconstruction has the advantage that one can unveil hidden noise elements like unwanted correlations using the entanglement entropy. This characteristic is tested to uncover non-local crosstalk that can explain partially why some pairs of qubits have low performance. For example, reconstructing the density state operator of nonconnected pairs of qubits (2,3) and (19,20) on ibm\_brisbane shows an entanglement entropy that deviates considerably from the mean entanglement entropy calculated for the interaction of (19,20) with all other pairs. It suggests that information leakage may occur at the multiplexed readout stage, a phenomenon that has been previously reported in \cite{Heinsoo_2018, bakr2024multiplexed, Maurya_2024}.

Finally, the ZECS shows that much more information about a quantum protocol on a given device can be recovered. It raises the question of how one can recover that information without having to reconstruct the whole density state operator, which quickly becomes impractical as the number of qubits increases. A future direction would be to understand what is the source of the noise removed and where exactly the remaining noise is coming.

\section*{Data Availability}
The datasets for problems used and/or analyzed during the current study are available from the following publicly accessible repository \url{https://github.com/alejomonbar/ZECS}.

\begin{acknowledgments}
\vspace{-10pt}

The authors thank Nick Bronn for the insightful discussions about some possible explanations of the non-local correlations observed. J. A. Montanez-Barrera acknowledges support from the German Federal Ministry of Education and Research (BMBF), the funding program Quantum Technologies - from basic research to market, project QSolid (Grant No. 13N16149). We acknowledge the use of IBM Quantum services for this work. The views expressed are those of the authors and do not reflect the official policy or position of IBM or the IBM Quantum team.
\end{acknowledgments}
\clearpage  % Start on a new page
\bibliography{References}
\clearpage  % Start on a new page
\onecolumngrid  % Switch to one column for the supplementary material
\appendix

\section{Supplementary Material}\label{Sec:appendix}
\subsection{Metrics}\label{Metrics}
The fidelity (F), trace distance (D), concurrence (C), and entanglement entropy ($S_{ab}$) are used to measure the quality of the solution given by CS and ZECS. The fidelity is expressed as 
\begin{equation}
    F = \Tr\left(\sqrt{\sqrt{\rho_1} \rho_2 \sqrt{\rho_1}}\right)^2, 
\end{equation}
where $\rho_1$ is the density state reconstruction and $\rho_2$ the ideal density state operator. The trace distance \cite{Nielsen2011} is defined by 
\begin{equation}
D(\rho_1, \rho_2) = \frac{1}{2}\mathrm{tr}|\rho_1 - \rho_2|,
\end{equation}
where $|A| = \sqrt{A^\dagger A}$ and $\rho_1$ and $\rho_2$ are the density state operators.

The concurrence \cite{Hill1997} is a metric of the entanglement of pairs of qubits and is given by 
\begin{equation}
    C(\rho) = \max \left( 0, \lambda_0 - \lambda_1 - \lambda_2 - \lambda_3 \right)
\end{equation}
where the $\lambda_i$ are the eigenvalues, sorted by magnitude from largest to smallest of the operator $R$ defined as
\begin{equation}
    R(\rho) = \sqrt{\sqrt{\rho}\bar{\rho}\sqrt{\rho}}
\end{equation}
where $\bar{\rho} = (\sigma_x \otimes \sigma_x) \rho^* (\sigma_x \otimes \sigma_x) $, $\sigma_x$ is the $x$-Pauli matrix, and $\rho^*$ is the complex conjugate of $\rho$.

\subsection{Zero Entropy}\label{Sec:ZE}
\begin{figure*}[!tbh]
\centering
\includegraphics[width=18cm]{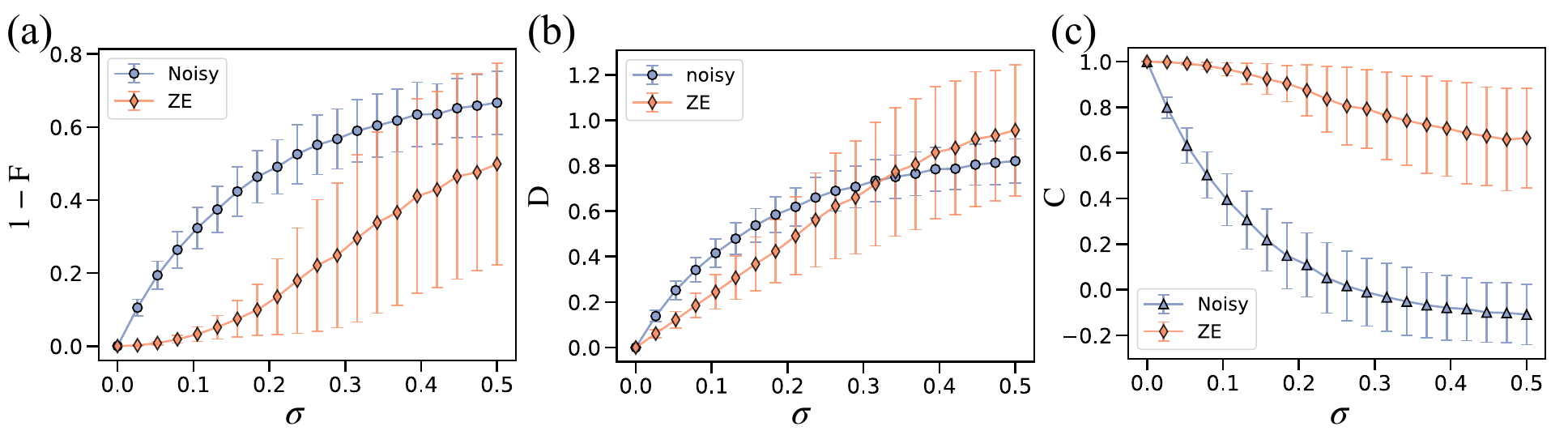}
\caption{\label{Fig:ZE} Bell state's perturbed density state operator compared to the reconstructed density state operator using ZE for different noise strengths in terms of the (a) infidelity, (b) trace distance, and (c) concurrence. The $x$-axis represents the standard deviation of the perturbation. The markers represent the mean value and the error bars represent the standard deviation of 1000 random perturbed states.}
\end{figure*}

The ZE methodology is general and can be applied to mixed states. In Fig.~\ref{Fig:ZE}, the capabilities of the ZE methodology to recover information from a noisy density state operator are shown. Random perturbations on a Bell state, $|\psi\rangle = \frac{1}{\sqrt{2}} \left(|00\rangle + |11\rangle\right)$ are used and  information is recovered using ZE. The random perturbation presented in \cite{Montanez-Barrera2022} is given by
\begin{equation}
\tilde\rho = \rho_0 + \frac{1}{2}\sum_{i,j} \eta_{ij} \sigma_i \otimes \sigma_j,
\end{equation}
and
\begin{equation}
\rho = \frac{(\sqrt{\tilde{\rho}})^\dagger \sqrt{\tilde{\rho}}}{\Tr(\tilde{\rho})} 
\end{equation}
where $\rho$ is a mixed state, $\rho_0 = |\psi\rangle \langle\psi|$, $\eta_{ij}$ is the random perturbation in the basis $i,j \in \{0,1,2,3\}$. $\eta_{ij}$ is taken randomly from a normal distribution with mean 0 and standard deviation, $\sigma$, varying from 0 to 0.5 to simulate different noise strengths.

The density operator $\rho_{ze}$ is first constructed from $\rho$ using Eqs. \ref{Eq:ze1} and \ref{Eq:ze2}. In Fig. \ref{Fig:ZE}, $\rho$ and $\rho_{ze}$ are compared in terms of the (a) infidelity ($1-F$), (b) $D$, and (c) $C$. These metrics are intended to give a sense of how close both states are to $\rho_0$. For $1-F$ and $D$, it is seen that for perturbations with $\sigma < 0.3$, the $\rho_{ze}$ is closer to the ideal $\rho_0$. In practice, this means that for QPUs with low error rates, ZE allows the reconstruction of a closer representation of $\rho_0$. In terms of $C$, which measures entanglement in pairs of qubits, using ZE, the entanglement remains stronger and for longer than the noisy case.

\begin{figure}[!tbh]
\centering
\includegraphics[width=8cm]{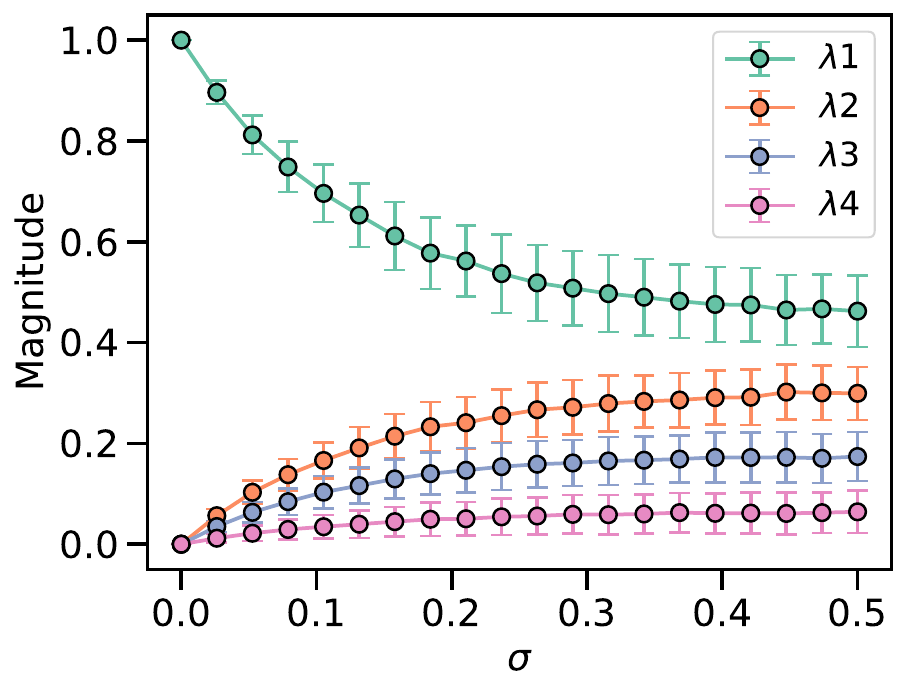}
\caption{\label{Fig:lambdas_RPE} Eigenvalues of the Bell state's perturbed density state operator for different noise strengths.}
\end{figure}

Figure \ref{Fig:lambdas_RPE} shows the 4 eigenvalues of the Bell state for the $ibm\_brisbane$ experiment (Fig. \ref{Fig:brisbane}(b)) when random perturbations are applied. As can be seen, as the strength of $\sigma$ increases, the information of the first eigenvalue decreases and the other eigenvalues begin to become more relevant. That the curves never cross as the strength of the noise increases indicates that the first eigenvalue can always be the largest of all the eigenvalues.

\subsection{ibm\_lagos extended analysis}\label{A:ibmlagos}

\begin{figure*}[t]
\centering
\includegraphics[width=17cm]{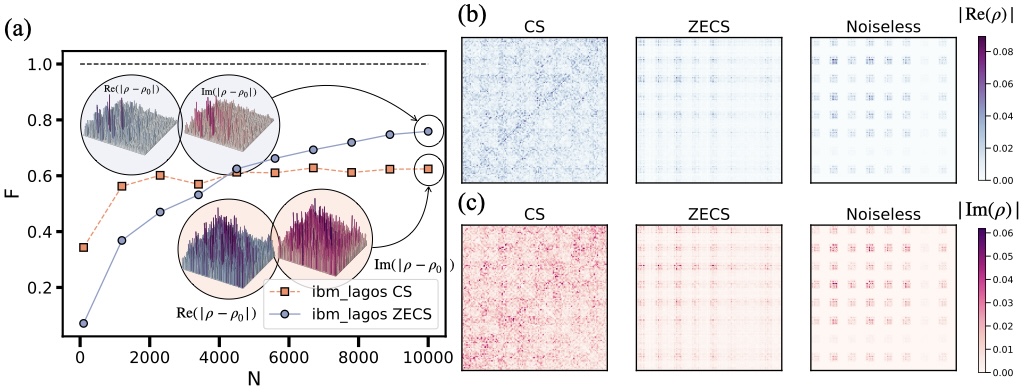}
\caption{\label{Fig:lagos_7q} Comparison between ZECS, CS, and a noiseless simulation of the density state operator reconstruction of the 7-qubits of ibm\_lagos for 4 repetitions of the protocol of Fig.~\ref{Fig:circuit_used}(a). (a) Fidelity versus the number of snapshots. The bars in the inset plots represent the noise in the real and imaginary components of the density state operator. Matrix elements of the (b) real and (c) imaginary components of the density state operator.}
\end{figure*}

Figure \ref{Fig:lagos_7q} provides the $F$ results for the reconstruction of the 7-qubit density state operator of ibm\_lagos using CS and ZECS. In particular, Fig. \ref{Fig:lagos_7q}(a) shows the results for $F$ versus the number of snapshots for 4 repetitions of the EfficientSU2 gate. Because of inherent noise on ibm\_lagos, after 2,500 snapshots, CS cannot improve the fidelity above 0.62. In contrast, applying ZECS, the fidelity improves up to 0.76. The results could improve even further using ZECS, but more snapshots would be needed. This is not the case for CS. The inset bar plots represent the relative error of the real and imaginary components of $\rho-\rho_0$.

Fig.\ref{Fig:lagos_7q}(b) and (c) show the real (blue) and imaginary (magenta) components of $\rho$ at $N=10,000$. Because of noise and limitations in sampling, the results deviate from the noiseless result. However, after applying ZECS to the CS information, part of the noise is removed so that the real and imaginary parts of the reconstructed density operator approach those for the noiseless simulation. In terms of fidelity, at the end of 10,000 snapshots, CS reaches a $F=0.756$ while ZECS reaches a $F=0.943$. ZECS is an assumption-free methodology and shows the information that a QPU can replicate from a quantum protocol. This is an advantage for diagnosis and characterization, allowing the focus to be on the inherent noise of the device instead of that associated with the instability of the device and the number of snapshots.

\subsection{ibm\_brisbane experiment}

Table \ref{tab:qubit_cases} summarizes the data for the EfficientSU2 experiments (two-qubit case), the EfficientSU2 2-qubit + idle qubit reconstruction (three-qubit case), and the pairs of 2-qubit reconstruction (four-qubit case) on ibm\_brisbane.

\begin{table*}[htbp]
\caption{\label{tab:qubit_cases} Summary of data for the two-, three-, and four-qubit cases involved in the reconstruction of the density state operators on ibm\_brisbane.}

\begin{tabular}{c|c|c|c|c|c|c|c|c|c|c}
\multicolumn{3}{c|}{\textbf{Two Qubit Cases}} & \multicolumn{4}{|c|}{\textbf{Three Qubit Cases}} & \multicolumn{4}{|c}{\textbf{Four Qubit Cases}} \\
\hline
\multicolumn{1}{|c|}{qubits} & $1 - F_{CS}$ & $1 - F_{ZECS}$ & qubits & $1 - F_{CS}$ & $1 - F_{ZECS}$ & $S_{ab}$ & qubits & $1 - F_{CS}$ & $1 - F_{ZECS}$ & \multicolumn{1}{|c|}{$S_{ab}$} \\
\hline
(13, 12) & 0.144 & 0.021 & (13, 12, 17) & 0.168 & 0.02 & 0.016 & (13, 12, 11, 10) & 0.401 & 0.132 & 0.099 \\
(11, 10) & 0.303 & 0.08 & (9, 8, 16) & 0.122 & 0.014 & 0.017 & (11, 10, 9, 8) & 0.355 & 0.09 & 0.107 \\
(9, 8) & 0.089 & 0.004 & (5, 4, 15) & 0.121 & 0.026 & 0.01 & (9, 8, 7, 6) & 0.211 & 0.041 & 0.086 \\
(7, 6) & 0.143 & 0.025 & (19, 20, 33) & 0.397 & 0.104 & 0.031 & (7, 6, 5, 4) & 0.243 & 0.069 & 0.08 \\
(5, 4) & 0.128 & 0.024 & (21, 22, 15) & 0.128 & 0.017 & 0.01 & (5, 4, 3, 2) & 0.308 & 0.078 & 0.063 \\
(3, 2) & 0.229 & 0.043 & (23, 24, 34) & 0.278 & 0.077 & 0.02 & (3, 2, 1, 0) & 0.339 & 0.058 & 0.102 \\
(1, 0) & 0.131 & 0.005 & (25, 26, 16) & 0.222 & 0.074 & 0.008 & (1, 0, 14, 18) & 0.271 & 0.043 & 0.074 \\
(14, 18) & 0.154 & 0.02 & (27, 28, 35) & 0.247 & 0.015 & 0.014 & (14, 18, 19, 20) & 0.472 & 0.143 & 0.117 \\
(19, 20) & 0.375 & 0.087 & (29, 30, 17) & 0.2 & 0.014 & 0.015 & (19, 20, 21, 22) & 0.442 & 0.126 & 0.145 \\
(21, 22) & 0.126 & 0.017 & (50, 49, 55) & 0.164 & 0.016 & 0.017 & (21, 22, 23, 24) & 0.353 & 0.097 & 0.072 \\
(23, 24) & 0.258 & 0.075 & (48, 47, 35) & 0.119 & 0.005 & 0.011 & (23, 24, 25, 26) & 0.42 & 0.193 & 0.115 \\
(25, 26) & 0.208 & 0.071 & (46, 45, 54) & 0.194 & 0.007 & 0.006 & (25, 26, 27, 28) & 0.373 & 0.107 & 0.058 \\
(27, 28) & 0.239 & 0.011 & (44, 43, 34) & 0.264 & 0.053 & 0.019 & (27, 28, 29, 30) & 0.376 & 0.035 & 0.081 \\
(29, 30) & 0.162 & 0.009 & (42, 41, 53) & 0.179 & 0.035 & 0.013 & (29, 30, 31, 32) & 0.292 & 0.069 & 0.101 \\
(31, 32) & 0.149 & 0.038 & (40, 39, 33) & 0.179 & 0.007 & 0.015 & (31, 32, 36, 51) & 0.254 & 0.089 & 0.049 \\
(36, 51) & 0.136 & 0.041 & (57, 58, 71) & 0.149 & 0.019 & 0.015 & (36, 51, 50, 49) & 0.288 & 0.07 & 0.098 \\
(50, 49) & 0.153 & 0.014 & (59, 60, 53) & 0.245 & 0.014 & 0.01 & (50, 49, 48, 47) & 0.239 & 0.028 & 0.088 \\
(48, 47) & 0.105 & 0.004 & (61, 62, 72) & 0.226 & 0.014 & 0.016 & (48, 47, 46, 45) & 0.265 & 0.033 & 0.081 \\
(46, 45) & 0.176 & 0.003 & (63, 64, 54) & 0.151 & 0.004 & 0.015 & (46, 45, 44, 43) & 0.356 & 0.084 & 0.074 \\
(44, 43) & 0.232 & 0.052 & (65, 66, 73) & 0.131 & 0.022 & 0.022 & (44, 43, 42, 41) & 0.342 & 0.113 & 0.151 \\
(42, 41) & 0.14 & 0.035 & (67, 68, 55) & 0.478 & 0.171 & 0.031 & (42, 41, 40, 39) & 0.327 & 0.05 & 0.075 \\
(40, 39) & 0.183 & 0.006 & (88, 87, 93) & 0.266 & 0.053 & 0.03 & (40, 39, 38, 37) & 0.288 & 0.03 & 0.08 \\
(38, 37) & 0.153 & 0.01 & (86, 85, 73) & 0.234 & 0.019 & 0.011 & (38, 37, 52, 56) & 0.26 & 0.044 & 0.066 \\
(52, 56) & 0.126 & 0.028 & (84, 83, 92) & 0.145 & 0.02 & 0.029 & (52, 56, 57, 58) & 0.254 & 0.057 & 0.074 \\
(57, 58) & 0.13 & 0.018 & (82, 81, 72) & 0.188 & 0.016 & 0.007 & (57, 58, 59, 60) & 0.324 & 0.061 & 0.078 \\
(59, 60) & 0.21 & 0.009 & (80, 79, 91) & 0.152 & 0.026 & 0.016 & (59, 60, 61, 62) & 0.339 & 0.052 & 0.171 \\
(61, 62) & 0.208 & 0.012 & (78, 77, 71) & 0.271 & 0.02 & 0.012 & (61, 62, 63, 64) & 0.293 & 0.026 & 0.09 \\
(63, 64) & 0.127 & 0.001 & (95, 96, 109) & 0.173 & 0.02 & 0.005 & (63, 64, 65, 66) & 0.243 & 0.039 & 0.047 \\
(65, 66) & 0.101 & 0.02 & (97, 98, 91) & 0.142 & 0.016 & 0.007 & (65, 66, 67, 68) & 0.517 & 0.19 & 0.078 \\
(67, 68) & 0.475 & 0.162 & (99, 100, 110) & 0.25 & 0.023 & 0.015 & (67, 68, 69, 70) & 0.603 & 0.222 & 0.134 \\
(69, 70) & 0.192 & 0.013 & (101, 102, 92) & 0.312 & 0.046 & 0.014 & (69, 70, 74, 89) & 0.374 & 0.061 & 0.105 \\
(74, 89) & 0.21 & 0.028 & (103, 104, 111) & 0.236 & 0.11 & 0.017 & (74, 89, 88, 87) & 0.383 & 0.102 & 0.05 \\
(88, 87) & 0.264 & 0.053 & (105, 106, 93) & 0.17 & 0.015 & 0.005 & (88, 87, 86, 85) & 0.43 & 0.095 & 0.099 \\
(86, 85) & 0.194 & 0.019 & (123, 122, 111) & 0.208 & 0.024 & 0.017 & (86, 85, 84, 83) & 0.299 & 0.045 & 0.08 \\
(84, 83) & 0.154 & 0.012 & (119, 118, 110) & 0.436 & 0.094 & 0.046 & (84, 83, 82, 81) & 0.256 & 0.041 & 0.075 \\
(82, 81) & 0.152 & 0.015 & (115, 114, 109) & 0.369 & 0.071 & 0.022 & (82, 81, 80, 79) & 0.256 & 0.042 & 0.068 \\
(80, 79) & 0.127 & 0.018 & & & & & (80, 79, 78, 77) & 0.36 & 0.061 & 0.101 \\
(78, 77) & 0.256 & 0.019 & & & & & (78, 77, 76, 75) & 0.399 & 0.138 & 0.121 \\
(76, 75) & 0.213 & 0.08 & & & & & (76, 75, 90, 94) & 0.348 & 0.115 & 0.066 \\
(90, 94) & 0.14 & 0.016 & & & & & (90, 94, 95, 96) & 0.274 & 0.054 & 0.103 \\
(95, 96) & 0.148 & 0.018 & & & & & (95, 96, 97, 98) & 0.205 & 0.041 & 0.051 \\
(97, 98) & 0.112 & 0.01 & & & & & (97, 98, 99, 100) & 0.332 & 0.053 & 0.123 \\
(99, 100) & 0.242 & 0.02 & & & & & (99, 100, 101, 102) & 0.433 & 0.104 & 0.109 \\
(101, 102) & 0.285 & 0.04 & & & & & (101, 102, 103, 104) & 0.428 & 0.141 & 0.096 \\
(103, 104) & 0.21 & 0.098 & & & & & (103, 104, 105, 106) & 0.321 & 0.142 & 0.209 \\
(105, 106) & 0.145 & 0.01 & & & & & (105, 106, 107, 108) & 0.28 & 0.038 & 0.055 \\
(107, 108) & 0.18 & 0.018 & & & & & (107, 108, 112, 126) & 0.284 & 0.065 & 0.129 \\
(112, 126) & 0.132 & 0.008 & & & & & (112, 126, 125, 124) & 0.406 & 0.072 & 0.056 \\
(125, 124) & 0.254 & 0.029 & & & & & (125, 124, 123, 122) & 0.383 & 0.079 & 0.086 \\
(123, 122) & 0.181 & 0.021 & & & & & (123, 122, 121, 120) & 0.362 & 0.077 & 0.175 \\
(121, 120) & 0.257 & 0.016 & & & & & (121, 120, 119, 118) & 0.568 & 0.171 & 0.196 \\
(119, 118) & 0.426 & 0.071 & & & & & (119, 118, 117, 116) & 0.515 & 0.159 & 0.105 \\
(117, 116) & 0.132 & 0.018 & & & & & (117, 116, 115, 114) & 0.455 & 0.119 & 0.1 \\
(115, 114) & 0.356 & 0.064 & & & & & & & & \\
\hline
\end{tabular}
\end{table*}

\end{document}